\documentclass[11pt]{article}
\usepackage{epsfig} 
\newcommand{\Aslash}{A \! \! \! /} 
\newcommand{\pslash}{p \! \! \! /} 
\newcommand{\partialslash}{\partial \! \! \! /} 
\newcommand{\half}{\mbox{\small{$\frac{1}{2}$}}} 
\newcommand{\Nc}{N_{\!c}} 
\newcommand{\Nf}{N_{\!f}} 
\newcommand{\MSbar}{\overline{\mbox{MS}}} 
\def \Tr {\mbox{Tr\,}}
\def \tr {\mbox{tr}}

\begin{document}
\hfill SPbU--IP--99--06

\hfill LTH--451 

{\begin{center}
{\LARGE {Quark mass anomalous dimension at $O(1/\Nf^2)$ in QCD} } \\ [8mm]
{\large M. Ciuchini$^a$, S.\'{E}. Derkachov$^b$\footnote{e-mail: 
Sergey.Derkachov@itp.uni-leipzig.de}, J.A. 
Gracey$^a$\footnote{e-mail: jag@amtp.liv.ac.uk} \& A.N. 
Manashov$^c$\footnote{e-mail: manashov@heps.phys.spbu.ru} \\ [3mm] } 
\end{center} 
} 

\begin{itemize}
\item[$^a$] Theoretical Physics Division, Department of Mathematical Sciences, 
\\
University of Liverpool, Liverpool, L69 7ZF, United Kingdom. 
\item[$^b$] Department of Mathematics, St Petersburg Technology Institute, \\
Sankt Petersburg, Russia \& \\ 
Institut f\"{u}r Theoretische Physik, Universit\"{a}t Leipzig, \\
Augustusplatz 10, D-04109 Leipzig, Germany. 
\item[$^c$] Department of Theoretical Physics, State University of St 
Petersburg, \\ 
Sankt Petersburg, 198904 Russia.
\end{itemize}

\vspace{3cm} 
\noindent 
{\bf Abstract.} We compute the $d$-dimensional critical exponents corresponding 
to the wave function and mass renormalization of the quark in QCD in the Landau
gauge at a new order, $O(1/\Nf^2)$, in the large $\Nf$ expansion. The 
computations are simplified by the establishment in $d$-dimensions of the 
critical point equivalence of QCD and the non-abelian Thirring model beyond
leading order. The form of the $O(1/\Nf^2)$ coefficients in the $\MSbar$
quark mass anomalous dimension at five loops is deduced and compared with the
numerical asymptotic Pad\'{e} approximant prediction.  

\newpage 

\setcounter{equation}{0}

The renormalization group equation, (RGE), plays a fundamental role in our 
understanding of the properties of quantum field theories. Central to this 
equation are the renormalization group functions such as the $\beta$-function
and the anomalous dimension, $\gamma(g)$, and for theories involving massive 
fields the mass anomalous dimension, $\gamma_m(g)$, also appears. Ordinarily 
such functions are computed order by order in the loop expansion. Though this 
clearly becomes more difficult at successive orders due to the increase in the
number of Feynman integrals which need to be computed. Despite this problem it 
has been possible to determine the $\beta$-function and quark mass anomalous 
dimensions in quantum chromodynamics, (QCD), at four loops in the $\MSbar$ 
renormalization scheme, [1-11]. Such calculations represent a significant 
achievement especially given that the order of $10^4$ Feynman diagrams have 
been evaluated at four loops to deduce the renormalization constants. To 
proceed further in perturbation theory will clearly be a colossal if not 
impossible undertaking. However, there do exist methods which probe the 
perturbative structure of the RG functions from another direction. Given that 
they depend not only on the coupling constant, $g$, but other parameters in the
theory such as the dimension of any internal symmetry groups, one can equally 
expand in one of these other variables. This will still involve the computation
of Feynman integrals but not those associated with ordinary perturbation 
theory. In QCD, with $\Nf$ flavours of quarks, they can be expanded in the 
large $\Nf$ expansion where $1/\Nf$ behaves as a bona fide perturbation 
parameter. Explicit details of the technique have been recorded in the 
literature but we emphasise that the extension of the $O(N)$ $\sigma$ model
methods of Vasil'ev et al, \cite{Vas,VN82}, to four dimensional gauge theories,
[14-18], offer the best and most efficient strategy to computing $1/\Nf$ 
information in QCD. For example, the anomalous dimension of the twist-$2$ 
operator dimensions which are fundamental to the operator product expansion 
used in deep inelastic scattering have been computed at $O(1/\Nf)$ and to all 
orders in the strong coupling constant, \cite{JGJB}. These results have been 
crucial in confirming the correctness of the explicit perturbative $3$-loop 
results of \cite{JMV97} which play a key role in the full two loop evolution of
the QCD structure functions. Clearly, given this important overlap with the 
current activity in explicit perturbative calculations, it is crucial that the 
large $\Nf$ method is developed to the next order, $O(1/\Nf^2)$. However, 
before such a deep inelastic programme can proceed, various fundamental 
computations need to be performed. In any ordinary renormalization the wave 
function renormalization constant is always computed first before determining 
the renormalization constants of the other parameters of the field theory. 
Likewise in the large $\Nf$ programme, the wave function critical exponent, 
$\eta$, must be deduced prior to the calculation of any other operator or field
dimension. Through the critical RGE, $\eta$ is related to $\gamma(g)$ by $\eta$
$=$ $2\gamma(g_{*})$ where $g_{*}$ is the value of the critical coupling at the 
$d$-dimensional fixed point of the QCD $\beta$-function.  

Whilst $\eta$ has been determined at $O(1/\Nf^2)$ in an arbitrary covariant 
gauge as a function of $d$ in QED, \cite{JG94}, the extension of that 
calculation to QCD has not yet been provided. This is one of the main aims of 
this letter where we will determine $\eta$ in a simpler model in the same 
universality class as QCD, known as the non-abelian Thirring model. As we will 
argue it has a simpler form than QCD since the triple and quartic gluon 
vertices which are present in Yang-Mills theories are absent but the expression
we obtain for $\eta$ will correspond to the quark wave function renormalization
constant in QCD itself. Clearly this reduces the number of Feynman diagrams 
which need to be considered. Indeed the $O(1/\Nf)$ calculations of say, 
\cite{JGJB}, were performed in the non-abelian Thirring model, (NATM). The 
connection with this model had previously been investigated at leading order in
the large $\Nf$ expansion in \cite{H2}. Although the provision of $\eta$ is 
fundamental to any future $O(1/\Nf^2)$ calculation, we have also computed the 
critical exponent which relates to the quark mass dimension, $\gamma_m(g_{*})$,
at $O(1/\Nf^2)$. There are several reasons for carrying out such a calculation.
First, $\eta$ is gauge dependent and is therefore not a fully meaningful 
physical quantity. On the contrary $\gamma_m(g_{*})$ is known to be a gauge 
independent (and scheme independent) exponent. This provides us with a 
non-trivial check on our computation, aside of course from the comparison with 
the explicit four loop $\MSbar$ results of \cite{VLR97,C97}. In addition, the 
Feynman integrals with a $[\bar{\psi}\psi]$ insertion are closely related to 
those required to deduce $\eta$ and so their computation does not in fact 
represent a significant amount of extra work. Moreover, as the calculation of 
$\gamma_m(g_{*})$ involves the insertion of the mass operator 
$[\bar{\psi}\psi]$ in a $2$-point quark Green function, it very much lays the 
foundation for the computation of the dimensions of similar operators which are
bilinear in the quark fields such as those which arise in deep inelastic 
scattering. 

The QCD Lagrangian in $d$ $=$ $4$ $-$ $2\epsilon$ dimensional Euclidean space 
reads
\begin{equation}
\label{LQCD}
S ~=~ \bar\psi^{iI}\not\!\!D \psi^{iI} ~+~ \frac{1}{4g^2} 
F^a_{\mu\nu}F^a_{\mu\nu} ~+~ \frac{1}{2\xi g^2}(\partial\cdot A)^2 ~+~ 
\partial_\mu \bar c^a \left(D^{\mu}c\right)^a ~,
\end{equation}
where $\psi^{iI}$ is the quark field belonging to the fundamental representation
of the colour group, $1$~$\leq$~$I$~$\leq$~$\Nf$, $A_\mu^a$ is the gluon field,
$c^a$ and ${\bar c}^a$ are the ghost fields in the adjoint representation of 
the colour group, $\xi$ is the covariant gauge parameter and $g$ is the 
coupling constant. The field strength tensor $F^a_{\mu\nu}$ and the covariant 
derivative are defined as $F_{\mu\nu}^a$ $=$ $\partial_\mu A_\nu^a$ $-$ 
$\partial_\nu A_\mu^a$ $+$ $f^{abc} A_\mu^a A_\nu^c$ and $D_\mu$ $=$ 
$(\partial_\mu$ $-$ $i A_\mu^a T^a)$, where $T^a$ are the group generators
in the corresponding representation and $f^{abc}$ are the structure constants
with $[T^a,T^b]$ $=$ $i f^{abc} T^c$. To ensure the coupling constant, 
$g$, is dimensionless below four dimensions we rescale it in the standard way 
by setting $g$ $\to$ $ M^{\epsilon}g$, where the parameter $M$ has dimensions 
of mass. 

The partition functions of the theory defined by the Lagrangian~(\ref{LQCD}) 
read
\begin{equation}
\label{corr}
\langle O_1(x_1)\ldots O_n(x_n)\rangle ~=~
Z^{-1}\int D\Phi\ O_1(x_1)\ldots O_n(x_n) \exp{\{-S\}} ~,
\end{equation}
where $\Phi\equiv\{A,\bar\psi,\psi,\bar c, c\}$ is the set of fundamental 
fields and $O_i(x_i)$ represent a basic field or a composite operator. The 
divergences arising in the calculation of~(\ref{corr}) are removed at each 
order of perturbation theory by the renormalization of the fields and 
parameters entering the QCD Lagrangian, as well as by renormalization of the 
composite operators. Hence the renormalized $1$-particle irreducible $n$-point 
Green functions with the insertion of $k$ multiplicatively renormalizable
composite operators satisfy the RGE 
\begin{equation}
\label{RG}
\left( M\partial_M+\beta_g\partial_g+\beta_\xi \partial_\xi
-n_\Phi \gamma_\Phi+\sum_{i=1}^k \gamma_{O_i} \right) \, 
\Gamma(x_1,\ldots,x_{n+k},M,g,\xi) ~=~ 0 ~,
\end{equation}
where $\gamma_{O_i}(g)$ is the anomalous dimension of the operator $O_i$, 
$n_\Phi\gamma_\Phi$ $=$ $n_A\gamma_A$ $+$ $n_\psi \gamma_\psi$ $+$ 
$n_c \gamma_c$ and the beta functions of the couplings $g$ and $\xi$ are given 
by $\beta_g$ $=$ $M\partial_M g$ and $\beta_\xi$ $=$ $M\partial_M \xi$. It 
should be noted here that the correlation functions of gauge invariant objects 
do not depend on the gauge fixing parameter, $\xi$, and in this case the term 
$\beta_\xi\partial_\xi$ drops out of Eq.~(\ref{RG}). 

Our analysis relies heavily on the existence of a non-trivial infra-red, (IR),
stable fixed point $g_{*}$ of the $d$-dimensional $\beta$-function, 
$\beta_g(g_{*})$ $=$ $0$, for large values of $\Nf$. The $\beta$-function has 
been calculated in $\MSbar$ using dimensional regularization and in the 
notation of the~\cite{PT84} with $a$ $=$ $(g/2\pi)^2$ 
is~\cite{GW73,CJ74,ET79,TVZ80,LV93} 
\begin{eqnarray}
\label{betag}
\beta_a(a)&=&(d-4)a ~+~ \left[\frac23 T_F \Nf-\frac{11}{6}C_A \right]a^2 ~+~
\left[\frac12 C_FT_F \Nf+\frac56 C_A T_F \Nf -\frac{17}{12}C_A^2 \right] a^3 
\nonumber\\
&& -~ \left[ \frac{11}{72} C_F T_F^2 \Nf^2 +\frac{79}{432} C_A T_F^2 \Nf^2+
\frac{1}{16} C_F^2 T_F \Nf -\frac{205}{288} C_FC_A T_F \Nf\right.
\nonumber \\
&& \left. ~~~~ -\frac{1415}{864} C_A^2 T_F \Nf +\frac{2857}{1728} C_A^3\right ] 
a^4 ~+~ O(a^5) ~,
\end{eqnarray}
from which it follows that
\begin{equation}
\label{gc}
a_{*} ~=~ \frac{3\epsilon}{T_F \Nf} ~+~ \frac{1}{4T_F^2 \Nf^2} \left( 
33C_A\epsilon-\left[ 27C_F+45C_A \right]\epsilon^2 + O(\epsilon^3) \right) ~+~ 
O\left( \frac{1}{\Nf^3} \right) ~.
\end{equation}
The Casimirs for a general classical Lie group are defined by 
\begin{equation}
\Tr \left(T^a T^b \right) ~=~ T_R \delta^{ab} ~, \ \ \ \ T^a T^a ~=~ C_F  I ~,
\ \ \ \ f^{acd} f^{bcd} ~=~ C_A \delta^{ab} ~.  
\end{equation}
It immediately follows from Eq.~(\ref{RG}) that the Green functions of gauge 
invariant operators are scale invariant at the critical point $g_{*}$. In other
words $G(\lambda x_i)$ $=$ $\lambda^{D_i} G(x_i)$, where $D_i$ is the scaling 
dimension of the corresponding Green function. Moreover, due to the IR nature 
of the fixed point, this index determines the power of the leading term of the 
IR asymptotic behaviour of the Green functions ($p_i\to 0$) near the critical 
points~\cite{Zinn}. On the contrary Green functions of gauge dependent objects, 
such as the propagators of the basic fields which will in general depend on 
$\xi$, are not invariant under scale transformations. Although one may 
restrict attention from the outset to gauge independent quantities, since they
have physical meaning, it is also possible and convenient to choose $\xi$ so 
that {\it all} Green functions are scale invariant. Evidently, this is 
equivalent to the condition $\beta_\xi(g_{*},\xi_{*})$ $=$ $0$. Since 
\begin{equation}
\beta_\xi (g, \xi)=-2\xi(\epsilon+\gamma_A+\beta_g/g) ~,
\end{equation} 
one concludes that the equation $\beta_\xi(g_{*},\xi_{*})=0$ has two solutions. 
One is $\xi_{*}$ $=$ $0$ whilst the other is $\gamma_A(g_{*},\xi_{*})$ $=$ $-$ 
$\epsilon$. Bearing in mind that our main aim is the development of the $1/\Nf$
expansion we choose the first solution, $\xi$ $=$ $0$, since the latter gives
$\xi$ $\sim$ $\Nf$, which leads to problems in the construction of the $1/\Nf$ 
scheme. The origin of the above two solutions for $\xi$ becomes more 
transparent if one tries to write down the most general form of the gluon 
propagator satisfying the requirements of both scale and gauge invariance. 
Indeed, scale invariance yields
\begin{equation}
G_{\mu\nu}(p) ~=~ \frac{M^{2\epsilon}}{p^{2\alpha}} \left(
A P_{\mu\nu}^\perp+B P_{\mu\nu}^\parallel\right) ~,
\end{equation}
where $P^{\perp}_{\mu\nu}$ and $P^\parallel_{\mu\nu}$ are the transverse and
longitudinal projectors, respectively, and $A$ and $B$ are constants. As is 
well known, \cite{Zinn}, radiative corrections do not contribute to the 
longitudinal part of gluon propagator. Hence, $G_{\mu\nu}^\parallel$ $=$
$\xi g^2 M^{2\epsilon} P_{\mu\nu}^\parallel p^{-2}$. This implies, that 
if $\alpha$~$\neq$~$1$ then $\xi$ must vanish, $\xi$ $=$ $0$. On the other 
hand for $\xi$ $\neq$ $0$ then one must have $\alpha$ $=$ $1$ which is easy to
check is equivalent to $\gamma_A$ $=$ $-$ $\epsilon$ corresponding to the 
canonical dimension of the field. Earlier work concerning the relation of 
scaling and conformal symmetry in the context of gauge theories has been given
in \cite{VPP}.  

It is well known from the theory of the critical phenomena~\cite{Zinn} that the
critical properties of the system do not depend on the details of the 
interactions but is determined mainly by ``global'' characteristics such as 
symmetries and the dimension of spacetime. Thus different systems may exhibit 
the same behaviour at the critical point. An example of this universality is the
fixed point relation between the Heisenberg ferromagnet and $\phi^4$ field 
theory. In what follows we construct the theory which belongs to the same 
universality class as QCD but which has a simpler structure. We first develop 
the $1/\Nf$ expansion for calculating correlators of the type given 
in~(\ref{corr}). This can be achieved in the standard manner by integrating 
over the fermion fields in the functional integral which  yields the following 
effective action for the gauge field,
\begin{equation}
\label{Seff}
S^{\mbox{eff}}_{A} ~\equiv~ \Nf\left (- \, 
\tr\ln(\!\not\!\partial-i\!\not\!\!A^{a}T^{a})
+\frac{M^{-2\epsilon}}{4\bar{g}^2} (F^a_{\mu\nu})^2 
+\frac{M^{-2\epsilon}}{2\xi\bar{g}^2}(\partial A)^2
\right) +\partial_\mu \bar c^a \left(D^{\mu}c\right)^a
\end{equation} 
where bearing in mind that $g_{*}^2$ $\sim$ $1/\Nf$ we have set $g^2$ $=$ 
$\bar g^2/\Nf$. The evaluation of the functional integral with 
action~(\ref{Seff}) by the saddle point method generates the systematic 
expansion for the correlators. If we now examine the IR asymptotic behaviour 
of the Green functions in this approach and first of all consider the gluon
propagator to first order in $1/\Nf$ in the Landau gauge, $\xi$~$=$~$0$, it is 
\begin{equation}
G_{\mu\nu}(p) ~=~ \Nf^{-1}\biggr[a p^{d-2} + 
\frac{M^{-2\epsilon}}{2\bar{g}^2}p^2\biggl]^{-1} P_{\mu\nu}^{\perp} ~,
\end{equation}
where the first term in the brackets arises from the fermion loop. It is 
obvious that the contribution coming from the $(F^a_{\mu\nu})^2$ term is less 
singular in comparison with the contribution of the fermion loop in the 
asymptotic limit $p$ $\to$ $0$. It can be easily checked that the diagrams with
vertices contained in $(F^a_{\mu\nu})^2$ also do not contribute to the leading 
order of the IR limit. So one concludes that this term does not influence the 
critical properties of the theory and according to the general scheme should be
excluded from action. Therefore we obtain the theory defined by the action
\begin{equation}
\label{NATM}
S ~=~ \bar\psi\,(\!\not\!\partial-i\!\not\!\!A^{a}T^{a})\,\psi ~+~ 
\frac{\Nf}{2\xi}(\sqcap\!\!\!\!\!\sqcup^{-\epsilon/2}\partial A)^2 ~+~ 
\partial_\mu \bar c^a \partial^\mu c ~+~  
f^{abc}\partial^\mu \bar c A_\mu^b c^c ~,
\end{equation}
which in the Landau gauge has the same critical behaviour as QCD. Of course, 
for gauge independent quantities it is true in any gauge. We have also modified
the form of the gauge fixing condition, in order that the transverse and 
longitudinal parts of the gluon propagator have the same momentum dependence. 
In fact, to derive~(\ref{NATM}) one can start from the theory with manifest 
gauge invariance which is determined by the action $S$ $=$ 
$\bar\psi( \partialslash - i\Aslash^{a}T^{a})\psi$ with ghost and gauge fixing 
terms in turn arising from the application of the Faddeev-Popov procedure to 
the functional integral. Power counting shows that~(\ref{NATM}) is
renormalizable within the $1/\Nf$ expansion. Of course, we assume that a gauge 
invariant regularization is used. The renormalized action then takes the form
\begin{equation}
\label{SR}
S_R ~=~ Z_1 \bar\psi\,\!\not\!\partial\psi ~-~
i Z_2\bar\psi\!\not\!\!A^{a}T^{a}\,\psi ~+~ 
\frac{\Nf}{2\xi}(\sqcap\!\!\!\!\!\sqcup^{-\epsilon/2}\partial A)^2 ~+~ 
Z_3\partial_\mu \bar c^a \partial^\mu c ~+~
Z_4f^{abc}\partial^\mu \bar c A_\mu^b c^c ~. 
\end{equation}
Due to the Slavnov-Taylor identities the renormalization constants $Z_i$ are
related by 
\begin{equation}
\label{ST}
Z_2\,Z_1^{-1} ~=~ Z_4\, Z_3^{-1} ~.
\end{equation} 
This was used in the exponent formulation to determine the ghost anomalous 
dimension at $O(1/\Nf)$ in \cite{JG93}. It should be noted that in the Landau 
gauge $Z_4$ $=$ $1$. As was proved above in the Landau gauge, the critical 
properties of QCD and this new theory which we shall refer to as the 
non-abelian Thirring model are identical. Therefore one can use the NATM model 
to deduce the QCD RG functions. This equivalence at leading order in $1/\Nf$ 
was noted in~\cite{H2} and used to deduce various exponents at $O(1/\Nf)$, 
\cite{JG91,JG92,JG94,JG93m,JGJB,JG93}. The extension of these calculations to 
$O(1/\Nf^2)$ requires special care. The main one is the necessity of using a 
gauge invariant regularization which was not crucial at $O(1/\Nf)$. The 
conventional dimensional regularization is not applicable here, since the gluon
propagator behaves as $p^{2-d}$ and the theory remains logarithmically 
divergent in any dimension $d$. To our knowledge most other invariant 
regularizations such as higher derivatives spoil the masslessness of the 
propagators, which makes higher order calculations virtually impossible. 
Usually in $1/N$ calculations the analytical regularization of~\cite{Vas} is 
used. However, this breaks gauge invariance.

We now consider how we can reconcile gauge invariance with the calculational
benefit of using massless propagators. First, we break gauge invariance 
of~(\ref{NATM}) from the beginning by introducing a new coupling, $\lambda$, 
for the ghost-gluon vertex in~(\ref{NATM}). For $\lambda$ $=$ $1$ one recovers 
the original model but the theory remains renormalizable for arbitrary 
$\lambda$ as well. The only effect will be that the identity~(\ref{ST}) will no 
longer hold. The bare coupling $\lambda_0$ is connected to the renormalized 
one, $\lambda$, by
\begin{equation}
\label{lam}
\lambda_0 ~=~ Z_\lambda \lambda ~=~ Z_4\,Z_1\,Z_2^{-1}Z_3^{-1} \lambda ~,
\end{equation}
where the $Z_i$ now depend on $\lambda$. Let us suppose that we used an
invariant method, such as regularization by higher derivatives~\cite{HD}, to 
regularize this extended theory. Then it immediately follows from~(\ref{ST})
and (\ref{lam}) that the equality $\lambda$ $=$ $1$ for the renormalized 
couplings implies that $\lambda_0$ $=$ $1$ as well. This leads us to the 
conclusion that $\lambda$ $=$ $1$ is a fixed point, $\beta_\lambda(1)$ $=$ $0$.
The existence of this fixed point is the key point and it does not depend on 
the regularization used. So using any other regularization can only change the 
position of the fixed point with in general $\lambda_{*}$ $=$ $1$ $+$ 
$O(1/\Nf)$. What is important, though, is that the anomalous dimensions 
calculated at the critical point, $\gamma(\lambda_{*})$, are scheme independent 
and, hence, coincide with the anomalous dimensions deduced in the original 
model~(\ref{NATM}). Therefore one can use the regularization which is most 
convenient from the computational point of view. Moreover, since we do not need
to consider diagrams with external ghost legs, then the only diagrams depending
on $\lambda$ are those with a ghost loop. As is evident from counting powers
of $1/\Nf$ these are themselves $O(1/\Nf^2)$. So at this order it is sufficient 
to set $\lambda$ $=$ $1$. 

In what follows we shall use the $\Delta$-regularization of~\cite{Vas,VN82}.
The propagators of the gluon, quark and ghost fields are obtained
from~(\ref{NATM}) as
\begin{equation}
G_{\nu\lambda}^{ab}(p) ~=~ \frac{\delta^{ab}}{n} 
\frac{G}{(p^2)^{\mu-1}}\left(P^{\perp}_{\nu\lambda}+
\tilde \xi P^\parallel_{\nu\lambda}\right ) ~, \ \ \ \ \  
D_\psi(p) ~=~ \frac{i\!\not\! p}{p^2} ~, \ \ \ \ \ 
D_c(p) ~=~ \frac{1}{p^2} ~, 
\end{equation} 
where we define $n$ to be the combination which arises naturally in the 
calculations as 
\begin{equation} 
n ~=~ \Nf \, T_F \, \mbox{Tr}_{\mbox{spinor}} \, I ~.
\end{equation} 
The value of the amplidude $G$ is derived from~(\ref{NATM}) as
\begin{equation}
G ~=~ (4\pi)^\mu\frac{\Gamma(2\mu)}{2\Gamma^2(\mu)\Gamma(2-\mu)} ~,
\end{equation} 
where we now set $\mu$ $\equiv$ $d/2$. The new gauge parameter $\tilde\xi$ is 
proportional to $\xi$ but in what follows we shall omit the tilde. The 
regularization of the theory is carried out by shifting the index of the gluon
propagator, by taking 
\begin{equation}
G_{\nu\lambda}^{ab}(p) ~=~ \frac{\delta^{ab}}{n}
\frac{GM^{2\Delta}}{(p^2)^{\mu-1+\Delta}}\left(P^{\perp}_{\nu\lambda}+
\xi P^\parallel_{\nu\lambda}\right ) ~,
\end{equation} 
where the factor $M^{2\Delta}$ is introduced to preserve the canonical 
dimension of the propagator. The divergences appearing in the diagrams as poles
in $\Delta$ are removed by the renormalization procedure. Throughout we will 
adopt the minimal subtraction scheme. The technical details and subtleties
which are inherent to this regularization can be found in the detailed 
discussion of~\cite{VN82,DM}. However, we record that up to $O(1/\Nf^2)$ there
exists a simple algorithm for the calculation of the anomalous 
dimensions,~\cite{DM}. To write down the corresponding formula in a compact 
form we introduce a factor $u$ for each the gluon propagator by setting 
\begin{equation} 
G_{\nu\lambda} ~\to~ u\, G_{\nu\lambda} ~, 
\end{equation} 
so that each diagram with $k$-internal gluon lines acquires a factor $u^k$. 
Then the anomalous dimensions of the basic fields $\Phi$ $=$ 
$\{\psi, A_\mu, c\}$ can be expressed via the renormalization constants 
$Z_\Phi$ where $\Phi_0$ $=$ $Z_\Phi \Phi$, as follows
\begin{equation}
\label{sf}
\gamma_\Phi ~=~ -~ \left. 2u\partial_u Z_\Phi^{(1)} \right|_{u=1} ~,
\end{equation} 
where 
\begin{equation}
Z_\Phi ~=~ 1 ~+~ Z^{(1)}_\Phi/\Delta ~+~ Z^{(2)}_\Phi/\Delta^2 ~+~ 
O \left( \frac{1}{\Delta^3} \right) ~. 
\end{equation} 
The matrix of anomalous dimensions of the system of composite operators,
$\{O_i\}$, which mix under renormalization, is given by
\begin{figure}[t]
\centerline{\epsfxsize12.0cm\epsfbox[100 250  500 400]{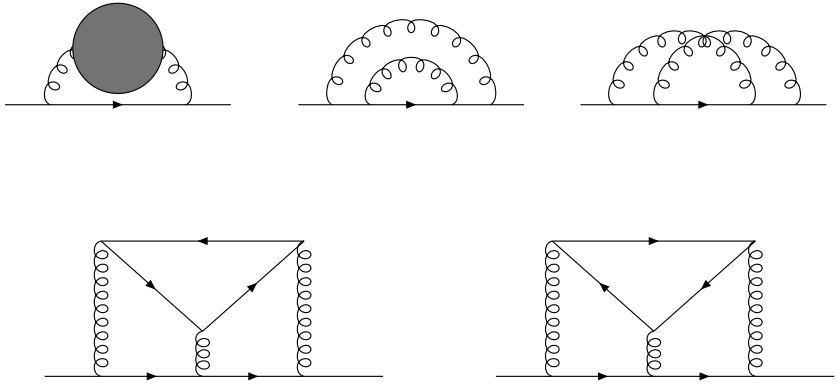}}
\caption[]{Diagrams contributing to the computation of $\eta_2$. The first
graph represents the gluon self energy diagrams of Fig.~\protect{\ref{fig2}.}} 
\label{fig1}
\end{figure}
\begin{equation}
\label{so}
\gamma_{ik} ~=~ \left. 2u\partial_u Z_{ik}^{(1)} \right|_{u=1} ~+~ 
\delta_{ik} n_{k,\Phi}\gamma_\Phi ~,
\end{equation}
where the mixing matrix, $Z_{ik}$, is defined in the standard way
\begin{equation} 
O_{i}^{R} ~=~ Z_{ik}O_{k} ~,
\end{equation} 
so that all Green functions of the operator $O_{i}^R$ are finite. Again, 
$Z^{(1)}_{ik}$ is the coefficient of the simple pole in $\Delta$. The 
derivation of~(\ref{sf}) and~(\ref{so}) can be found in~\cite{DM}. Though it 
should be stressed that these formul{\ae} are valid only up to $O(1/\Nf^2)$. 
We note also the obvious resemblance of the formul{\ae}~(\ref{sf}) 
and~(\ref{so}) with those used in dimensional regularization. 

Before proceeding to the results we underline an advantage of the approach of
\cite{VN82,DM} in comparison with the method of the self-consistency equations,
(SE), of~\cite{Vas,JG94} which has been more widely used in the $1/N$ 
computations. In the SE method to find the critical exponent $\eta$ one has to 
calculate the corresponding renormalized Green functions and then solve the 
self-consistency equations. By contrast to calculate critical exponents using 
(\ref{sf}) and~(\ref{so}) one only needs to know the {\it divergent} part of 
the corresponding diagrams. This is much simpler from a computational point of 
view. As is well known progress in multiloop perturbative calculations relies 
heavily on the possibility of expressing the RG functions through the 
renormalization constants. For calculations of the latter there are various
calculational shortcuts. We now demonstrate the effectiveness of this approach 
by calculating the exponent $\eta$ $=$ $2\gamma(g_{*})$ and the anomalous 
dimensions of the $[\bar\psi\psi]$ operator.

\begin{figure}[t]
\centerline{\epsfxsize10.0cm\epsfbox[100 150  500 400]{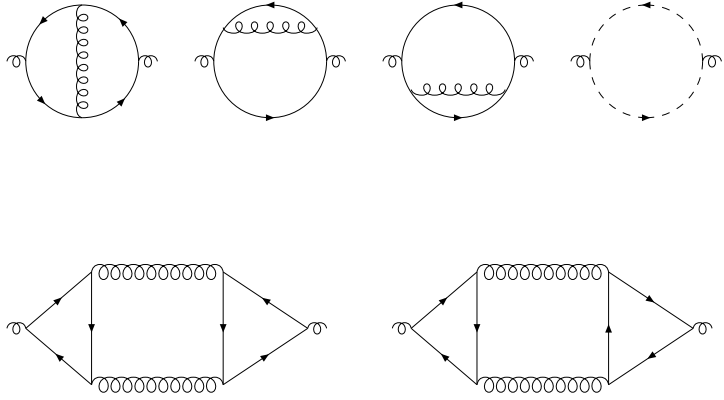}}
\caption[]{The diagrams contributing to the gluon self-energy at $O(1/\Nf^2)$.}
\label{fig2}
\end{figure}
The diagrams which contribute to $Z_\psi$ at $O(1/\Nf^2)$ are illustrated in
Figs.~\ref{fig1} and~\ref{fig2} where in the latter the dashed line corresponds
to the ghost field. We now take $\lambda$ $=$ $1$. Having calculated the 
contributions to $Z_\psi$ from the diagrams in both figures, we find the 
following expression for $\eta$ $=$ $2\gamma(g_{*})$ at $O(1/\Nf^2)$. If we set 
\begin{eqnarray}
\label{eta}
\eta &=& \eta_1/n ~+~ \eta_2/n^2 ~+~ O(1/n^3) ~, \\
\label{eta1}
\eta_1 &=& C_F\,\eta_0 ~, \\
\label{eta2}
\eta_2 &=& C_F^2\, \eta_a ~+~ C_F\, C_A\,\eta_b ~,
\end{eqnarray}
where
\begin{equation}
\eta_{0} ~=~ \frac{(\mu-2)(2\mu-1)\Gamma(2\mu)} 
{\Gamma^2(\mu)\Gamma(\mu+1)\Gamma(2-\mu)}
\end{equation} 
then in the Landau gauge  
\begin{equation} 
\eta_a ~=~ \frac{(\mu-1)\eta_0^2}{(\mu-2)(2\mu-1)}\left [
\frac{2(\mu-1)(\mu-3)}{\mu(\mu-2)} ~+~ 3\mu\left( \Theta-\frac{1}{(\mu-1)^2}
\right) \right] 
\end{equation} 
which has been computed previously in \cite{JG94}, and 
\begin{eqnarray} 
\eta_b &=& \eta_0^2\left [ 
\frac{(12\mu^4-72\mu^3+126\mu^2-75\mu+11)}{2(2\mu-1)^2(2\mu-3)(\mu-2)^2} ~-~ 
\frac{\mu(\mu-1)}{2(2\mu-1)(\mu-2)} \left( \Psi^2+\Phi \right) \right. 
\nonumber \\ 
&& \left. ~~~~+~ \frac{(8\mu^5-92\mu^4+270\mu^3-301\mu^2+124\mu-12)\Psi} 
{4(2\mu-1)^2(2\mu-3)(\mu-2)^2} \right ] ~.
\end{eqnarray}
The functions $\Psi$, $\Phi$ and $\Theta$ are defined as
\begin{eqnarray}
\Psi(\mu) &=& \psi(2\mu-3) ~+~ \psi(3-\mu) ~-~ \psi(1) ~-~ \psi(\mu-1) ~, 
\nonumber\\
\Phi(\mu) &=& \psi^\prime(2\mu-3) ~-~ \psi^\prime(3-\mu) ~-~ 
\psi^\prime(\mu-1) ~+~ \psi^\prime(1) ~, \\
\Theta(\mu) &=& \psi^\prime(\mu-1) ~-~ \psi^\prime(1) ~, \nonumber
\end{eqnarray}
where $\psi(x)$ $=$ $(\ln{\Gamma(x)})^\prime$. The technical details of the 
calculations and the values for the individual graphs will be given 
elsewhere~\cite{CDGM}. 

Next to determine the quark mass anomalous dimension at $O(1/\Nf^2)$ one 
computes the anomalous dimension of the associated composite operator
$[\bar\psi\psi]$. At $O(1/\Nf)$ this was derived in~\cite{JG93} as well as the 
anomalous dimensions of the basic fields. The diagrams contributing to its 
renormalization constant, $Z_{\bar\psi\psi}$, are obtained from those for the 
quark propagator of Figs.~\ref{fig1} and \ref{fig2} by the insertion of the
operator in the fermion lines connected to external vertices. In fact, it is 
sufficient to calculate the diagrams arising from the two $2$-loop diagrams in 
the first line of Fig.~\ref{fig1}. The contributions to $Z_{\bar\psi\psi}$ 
from the other diagrams can be related to the contributions of the 
corresponding diagrams to $Z_{\psi}$ by  
\begin{equation}
\label{etamass}
\delta Z_{\bar\psi\psi} ~=~ -~ \frac{2\mu}{(\mu-2)}\cdot\delta Z_{\psi} ~~ 
\Rightarrow ~~ \delta\gamma_{m} ~=~ -~ \frac{2}{(\mu-2)}\cdot\delta\eta_2 
\end{equation} 
and 
\begin{equation} 
\gamma_m ~=~ \eta ~+~ \gamma_{\bar\psi\psi} ~.
\end{equation}
This follows if we take the flow of the external momenta $p$ in the propagator
diagrams along the fermion lines and differentiate the corresponding integrals
with respect to $p_\mu$. Since 
\begin{equation} 
\frac{\partial}{\partial p^\mu} \, \frac{\pslash}{p^2} ~=~ -~ 
\frac{\pslash \gamma^\mu \pslash}{(p^2)^2} 
\end{equation} 
the resulting diagrams will have the same topology as those for the 
$[\bar\psi\psi]$ operator with the only difference being the presence of a new
insertion containing $\gamma^\mu$. Further, we use the fact that the pole part 
of the diagrams after the subtraction of the divergent subgraphs does not 
depend on the external momenta. So we choose the momenta flow as shown in 
\begin{figure}[t]
\centerline{\epsfxsize12.0cm\epsfbox[200 180  500 300]{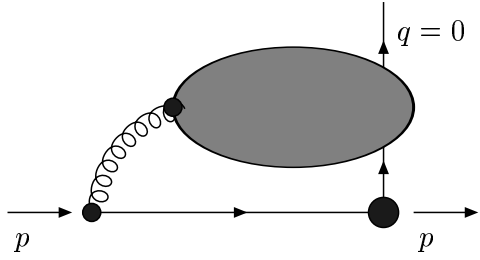}}
\caption[]{External momenta routing in the quark $2$-point function with an  
operator insertion.}
\label{fig3}
\end{figure}
Fig.~\ref{fig3}. The big black dot in Fig.~\ref{fig3} denotes the insertion of 
the unit matrix $I$ for the mass operator diagrams, and $\gamma_\mu$ for the 
propagator ones. Since the insertion of $I$ or $\gamma_\mu$ in the vertex 
influences only the $\gamma$ matrix structure of the diagram and does not touch
the momentum integral, it can be easily checked that the 
identity~(\ref{etamass}) holds irrespective of the explicit structure of the 
coloured block on the Fig.~\ref{fig3}. Eventually, using~(\ref{so}) we find for
$\gamma_m(g_{*})$
\begin{equation}
\gamma_m(g_{*}) ~=~ \gamma_{m,1}/n ~+~ \gamma_{m,2}/n^2 ~+~ O(1/n^3) ~,
\end{equation}
where
\begin{eqnarray}
\gamma_{m,1} &=& -~ \frac{2\,C_F\eta_0}{(\mu-2)} \\
\gamma_{m,2} &=& C_F^2 \gamma_a ~+~ C_F C_A \gamma_b ~,
\end{eqnarray}
and $\gamma_a$ and $\gamma_b$ are given by
\begin{eqnarray}
\gamma_a &=& -~ \frac{2\eta_a}{(\mu-2)} ~-~ \frac{2(2\mu^2-4\mu+1)\eta_0^2} 
{(\mu-2)^3(2\mu-1)} \\
&&\nonumber \\
\gamma_b &=& -~ \frac{2\eta_b}{(\mu-2)} ~+~ \frac{\mu^2(2\mu-3)^2\eta_0^2} 
{4(\mu-2)^3(\mu-1)(2\mu-1)} ~.
\end{eqnarray}
where the former was determined in \cite{JG93m}. Having derived a 
$d$-dimensional expression for the quark mass dimension, we make several 
remarks on its properties. First, if one sets $\mu$ $=$ $2$ $-$ $\epsilon$ and 
expands in powers of $\epsilon$, the coefficients to and including 
$O(\epsilon^4)$ agree with those determined from the explicit $\MSbar$ four 
loop perturbative mass dimension of \cite{NRT79,TNW81,T82,VLR97,C97}. Given 
that we have only evaluated three loop diagrams agreement at this order 
represents a non-trivial check on our analysis. In addition we have also 
computed the exponent in an arbitary covariant gauge and checked that in the 
final sum the gauge parameter cancels. These details are presented in 
\cite{CDGM}. Therefore, we are confident that our value for $\gamma_{m,2}$ is 
correct. This allows us now to produce new information on the mass anomalous 
dimension at $5$-loops in $\MSbar$. First, we write the $O(1/\Nf^2)$ form of 
the mass anomalous dimension as 
\begin{eqnarray} 
\gamma_m(a) &=& -~ \frac{3}{2} C_F a ~+~ \left( \frac{10}{24} T_F \Nf 
- \frac{3}{16} C_F - \frac{97}{48} C_A \right) C_F a^2 \nonumber \\
&& +~ \sum_{r=3}^\infty \left( m_{r0} T_F^{r-1} \Nf^{r-1} + m_{r1} T_F^{r-2}  
\Nf^{r-2} \right) C_F a^r ~+~ O \left( \frac{1}{\Nf^3} \right) ~,  
\end{eqnarray} 
where the order symbol means that we are ignoring contributions from 
$\gamma_{m,3}$ and six loop terms. To extract the coefficients $m_{50}$ and 
$m_{51}$ requires $a_*$ at $O(\epsilon^5)$ and $O(1/\Nf^2)$. Although the full
$5$-loop $\MSbar$ QCD $\beta$-function is not yet available, the critical
coupling which we require at $O(1/\Nf^2)$ has already been computed in 
\cite{JG96}. Hence, we find that 
\begin{eqnarray} 
m_{50} &=& \frac{5\zeta(3)}{162} ~-~ \frac{\zeta(4)}{18} ~+~ \frac{65}{2592} \\ 
m_{51} &=& \left[ \frac{5\zeta(4)}{8} ~-~ \frac{\zeta(5)}{3} ~-~ 
\frac{11\zeta(3)}{48} ~-~ \frac{4483}{20736} \right] C_F \nonumber \\
&& +~ \left[ \frac{8\zeta(5)}{9} ~-~ \frac{17\zeta(4)}{36} ~-~ 
\frac{671\zeta(3)}{1296} ~-~ \frac{18667}{124416} \right] C_A  
\label{masscoeff} 
\end{eqnarray} 
where $\zeta(n)$ is the Riemann zeta function. 

Whilst this gives an indication of the exact form of these coefficients at 
$5$-loops, we have another motivation for computing it. Recently, there has 
been activity in trying to estimate the higher order coefficients of 
perturbative functions in four dimensional field theories from knowledge of the
known lowest orders and the asymptotic behaviour at high orders, 
\cite{SEK95,EKS97}. This technique known generally as the asymptotic Pad\'{e} 
approximant method, (APAP), has had varying degree of success. For instance, 
the $4$-loop $\beta$-function of ${\cal N}$ $=$ $1$ supersymmetric QCD had been
determined analytically in \cite{JJS97} up to one unknown parameter. By 
applying APAP methods and its refinement, WAPAP, this parameter was determined 
numerically. Subsequent explicit calculations in \cite{JJP98} produced this 
parameter analytically and it was in good agreement with the APAP estimate for 
it. On the other hand the original application of the APAP method to the 
$4$-loop $\MSbar$ QCD $\beta$-function, \cite{EKS97}, did not yield as accurate 
a prediction. In this case the coefficients of the polynomial in $\Nf$ which 
appears at $4$-loops were estimated with the leading coefficient fixed from the
known result of the $1/\Nf$ expansion, \cite{JG96}. One of the reasons for a 
less accurate prediction was the appearance of new colour group Casimirs at 
four loops which were absent in lower orders and which therefore had not been 
built into the estimating procedure. Taking account of these issues the method 
was refined in \cite{EJJKS98} when the full $4$-loop $\MSbar$ result became 
available and new estimates were provided instead for the $5$-loop 
$\beta$-function and quark mass dimension. In particular the coefficients of 
the $\Nf$-polynomials were estimated. Whilst we believe it will be some time 
before the full five loop $\MSbar$ renormalization of QCD will be performed, if
the APAP method for higher order estimation is to serve any useful purpose in 
the interim it is important to determine its reliability. The determination of 
(\ref{masscoeff}) for QCD therefore provides us with that test. In 
\cite{EJJKS98} the coefficient of the dominant $\Nf$ term in the $5$-loop $\Nf$
polynomial was fixed by the leading order large $\Nf$ coefficient, $m_{50}$. 
The results were quoted for various values of $\Nc$ for two cases. One was 
where the effect of the new $4$-loop colour Casimirs, $Q_4$, was accounted for 
and the other was the case where their presence was ignored. We have reproduced
the values of these coefficients for the choices of $\Nc$ given in 
\cite{EJJKS98} together with the numerical values of (\ref{masscoeff}). In 
order to facilitate comparison we note the relation between our notation and 
that of \cite{EJJKS98} is
\begin{equation} 
E_4 ~=~ -~ \half T_F^4 \, C_F \, m_{50} ~~,~~ D_4 ~=~ -~ 
\half T_F^3 \, C_F \, m_{51} ~.  
\end{equation} 
\begin{table}[hb] 
\begin{center} 
\begin{tabular}{|c||c|c|c|} 
\hline 
$N_{\!c}$ & $D_4$ (w $Q_4$) &  
$D_4$ (w/o $Q_4$) & 
$D_4^{\mbox{\footnotesize{exact}}}$ \\  
\hline 
 2 & $8.12 \times 10^{-3}$ & $8.88 \times 10^{-3}$ & $0.0396$ \\ 
 3 & $0.037$ & $0.037$ & $0.1083$ \\ 
 4 & $0.0891$ & $0.0831$ & $0.2049$ \\ 
 5 & $0.165$ & $0.148$ & $0.3292$ \\ 
20 & $4.31$ & $3.48$ & $5.5113$ \\  
\hline 
\end{tabular} 
\end{center} 
\begin{center} 
{Table 1. Comparison of APAP results for $D_4$ with the numerical value of the 
exact coefficient.}  
\end{center} 
\end{table} 
{}From the table it would appear that the WAPAP estimates are not in agreement 
with (\ref{masscoeff}). However, it is worth putting the estimates in the 
context of the other coefficients. For $\Nc$ $=$ $3$, for instance, the leading 
coefficient of the $\Nf$ polynomial is $O(10^{-5})$ whilst the constant term 
has a WAPAP prediction of $530$ which is roughly seven orders of magnitude 
larger. That the WAPAP prediction for $m_{51}$ is within an order of magnitude
is perhaps remarkable particularly given the nature of the estimation method.
By fitting to the large order asymptotic behaviour of the perturbation series
one is essentially ensuring that the constant term of the polynomial in $\Nf$ 
is close to the correct value in the WAPAP as this will always be the dominant 
contribution for a range of values of $\Nf$ where $\Nf$ is relatively small. 
Therefore, one would not fully expect that the first few coefficients to be 
reliably predicted. Moreover, in light of our exact evaluation of $m_{51}$, it 
would seem worthwhile to refine the WAPAP estimate of the other coefficients by
taking our value either as an extra normalization or another constraining 
number to fit to. 

To conclude our article, we emphasise that we have computed the critical 
exponents corresponding to the wave function and mass anomalous dimension 
in $d$-dimensions at a new order in the $1/\Nf$ expansion in QCD. Not only have
we provided new information on the perturbative structure of these RG functions
but we have also demonstrated the viability of the large $\Nf$ procedure to 
compute information on the other important quantities such as those which 
relate to operators in deep inelastic scattering. 

\vspace{1cm}  
\noindent 
{\bf Acknowledgements.} This work was carried out with the support of the
Russian Fond of Basic Research, Grant 97--01--01152 (SED and ANM), in part by 
DFG Project N KI-623/1-2 (SED), a University of Pisa Exchange Research 
Fellowship (MC) and from PPARC through an Advanced Fellowship (JAG). Also SED 
and ANM would like to thank Prof.~A.N.~Vasil'ev for useful discussions. 
Invaluable to the calculations were the symbolic manipulation programme 
{\sc Form} and computer algebra package {\sc Reduce}.


\newpage 
\begin{thebibliography}{99}
\bibitem{GW73} D.J.\ Gross \& F.J.\ Wilczek, Phys.\ Rev.\ Lett. {\bf 30} (1973)
1343; \\
H.D.\ Politzer, Phys.\ Rev.\ Lett. {\bf 30} (1973) 1346.
\bibitem{CJ74} W.E.\ Caswell, Phys.\ Rev.\ Lett. {\bf 33} (1974) 244;\\
D.R.T.\ Jones,  Nucl.\ Phys. {\bf B75} (1974) 531.
\bibitem{ET79} E.S.\ Egorian \& O.V.\ Tarasov, Theor.\ Math.\ Phys. {\bf 41} 
(1979) 26.
\bibitem{TVZ80} O.V.\ Tarasov, A.A.\ Vladimirov \& A.Yu.\ Zharkov, Phys.\ Lett.
{\bf B93} (1980) 429.
\bibitem{LV93} S.A.\ Larin \& J.A.M.\ Vermaseren, Phys.\ Lett. {\bf B303} 
(1993) 334.
\bibitem{RVL97} T.\ van Ritbergen, J.A.M.\ Vermaseren \& S.A.\ Larin, Phys.\ 
Lett.\ {\bf B400} (1997), 379. 
\bibitem{NRT79} D.V.\ Nanopoulos \& D.A.\ Ross, Nucl.\ Phys.\ {\bf B157} (1979)
273. 
\bibitem{TNW81} R.\ Tarrach, Nucl.\ Phys.\ {\bf B183} (1981) 384; \\ 
O.\ Nachtmann \& W.\ Wetzel, Nucl.\ Phys.\ {\bf B187} (1981) 333. 
\bibitem{T82} O.\ Tarasov, JINR preprint P2-82-900.  
\bibitem{VLR97} J.A.M.\ Vermaseren, S.A.\ Larin \& T.\ van Ritbergen, Phys.\ 
Lett.\ {\bf B405} (1997) 327.  
\bibitem{C97} K.G.\ Chetyrkin, Phys.\ Lett.\ {\bf B404} (1997) 161.  
\bibitem{Vas} A.N.\ Vasil'ev, Yu.M.\ Pis'mak \& J.R.\ Honkonen, Theor.\ Math.\ 
Phys. {\bf 46} (1981) 157; Theor.\ Math.\ Phys. {\bf 47} (1981) 291.  
\bibitem{VN82} A.N.\ Vasil'ev \& M.Yu.\ Nalimov, Theor.\ Math.\ Phys. {\bf 56} 
(1982) 643.
\bibitem{JG91} J.A.\ Gracey, J.\ Phys.\ {\bf A24} (1991) L431.
\bibitem{JG92} J.A.\ Gracey, Int.\ J.\ Mod.\ Phys.\ {\bf A8} (1993) 2465.
\bibitem{JG94} J.A.\ Gracey, Nucl.\ Phys.\ {\bf B414} (1994) 614. 
\bibitem{JG93m} J.A.\ Gracey, Phys.\ Lett.\ {\bf B317} (1993) 415.
\bibitem{JG96} J.A.\ Gracey, Phys.\ Lett.\ {\bf B373} (1996) 178.
\bibitem{JGJB} J.A.\ Gracey, Phys.\ Lett.\ {\bf B322} (1994) 141; \\ 
J.F.\ Bennett \& J.A.\ Gracey, Nucl.\ Phys.\ {\bf B517} (1998) 241.   
\bibitem{JMV97} S.A.\ Larin, P.\ Nogueira, T.\ van Ritbergen \& J.A.M.\ 
Vermaseren, Nucl.\ Phys.\ {\bf B492} (1997) 338. 
\bibitem{H2} A.\ Hasenfratz \& P.\ Hasenfratz, Phys.\ Lett. {\bf B297} (1992) 
166.
\bibitem{PT84} P.\ Pascual \& R.\ Tarrach, {\it  QCD: Renormalization for the 
Practioner} Lecture Notes in Physics 194 (Springer, Berlin, 1984).
\bibitem{Zinn} J.\ Zinn-Justin, {\it Quantum Field Theory and Critical 
Phenomena} (Clarendon, Oxford, 1990).  
\bibitem{VPP} A.N.\ Vasil'ev, M.M.\ Perekalin \& Yu.M.\ Pis'mak, Theor.\ Math.\
Phys.\ {\bf 55} (1982) 323; Theor.\ Math.\ Phys.\ {\bf 60} (1984) 317.  
\bibitem{JG93} J.A.\ Gracey, Phys.\ Lett.\ {\bf B318} (1993) 177.
\bibitem{HD} A.A.\ Slavnov, Nucl.\ Phys.\ {\bf B31} (1971) 301. 
\bibitem{DM} S.\'{E}.\ Derkachov, A.N.\ Manashov, Nucl.\ Phys. {\bf B522} 
(1998) 301; Teor.\ Math.\ Phys. {\bf 116} (1998) 379.
\bibitem{CDGM} M.\ Ciuchini, S.\'{E}.\ Derkachov, J.A.\ Gracey \& A.N.\ 
Manashov, {\it in preparation}. 
\bibitem{SEK95} M.A.\ Samuel, J.\ Ellis \& M.\ Karliner, Phys.\ Rev.\ Lett.\
{\bf 74} (1995) 4380. 
\bibitem{EKS97} J.\ Ellis, M.\ Karliner \& M.A.\ Samuel, Phys.\ Lett.\ {\bf 
B400} (1997) 176. 
\bibitem{JJS97} I.\ Jack, D.R.T.\ Jones \& M.A. Samuel, Phys.\ Lett.\ {\bf 
B407} (1997) 143.  
\bibitem{JJP98} I.\ Jack, D.R.T.\ Jones \& A.\ Pickering, Phys.\ Lett.\ {\bf
B435} (1998) 61. 
\bibitem{EJJKS98} J.\ Ellis, I.\ Jack, D.R.T.\ Jones, M.\ Karliner \& M.A.\
Samuel, Phys.\ Rev.\ {\bf D57} (1998) 2665. 
\end{thebibliography}
\end{document}